\documentclass[conference]{IEEEtran}
\IEEEoverridecommandlockouts
\usepackage[letterpaper, portrait, margin=0.7in, bmargin=1in]{geometry}

\usepackage{cite}
\usepackage{amsmath,amssymb,amsfonts}
\usepackage{algorithmic}
\usepackage{graphicx}
\usepackage{textcomp}
\usepackage{xcolor}
\usepackage[cmintegrals]{newtxmath}
\usepackage{bm}
\usepackage{enumitem}
\usepackage{dblfloatfix}
\usepackage{ragged2e} 
\usepackage{adjustbox}

\usepackage{gensymb}
\usepackage{ragged2e}

\usepackage{tabularx}
\usepackage{makecell}

\usepackage{array}
\usepackage{multirow}

\usepackage{subfig}
\usepackage{subcaption}

\usepackage{adjustbox} 
\usepackage{booktabs}
\usepackage{colortbl} 
\usepackage{xcolor} 

\usepackage[font=footnotesize]{caption}

\newcolumntype{?}{!{\vrule width 2pt}}

\definecolor{lightgray}{gray}{0.95}

\def\BibTeX{{\rm B\kern-.05em{\sc i\kern-.025em b}\kern-.08em
    T\kern-.1667em\lower.7ex\hbox{E}\kern-.125emX}}

\usepackage{fancyhdr}
\fancypagestyle{firststyle}{\fancyhf{}
		\fancyhead[L]{D. Shakya, N. A. Abbasi, M. Ying, I. Jariwala, J. J. Qin, I. S. Gupte, B. Meier, G. Qian, D. Abraham, T. S. Rappaport, A. F. Molisch,  ``Standardized Machine-Readable Point-Data Format for Consolidating Wireless Propagation Across Environments, Frequencies, and Institutions",\textit{(accepted) IEEE MILCOM 2025, }Los Angeles, USA, Oct. 2023, pp. 1--6.}
	}
\usepackage{times}

\begin{document}

 \title{Standardized Machine-Readable Point-Data Format for Consolidating Wireless Propagation Across Environments, Frequencies, and Institutions
 \thanks{The work at NYU was supported by NYU WIRELESS Industrial Affiliates Program. The work at USC was supported by Samsung Research America and the NSF under grants 2229535, 2133655.}}

\author{\IEEEauthorblockN{Dipankar Shakya\IEEEauthorrefmark{1}, Naveed A. Abbasi\IEEEauthorrefmark{2}, Mingjun Ying\IEEEauthorrefmark{1}, 
Isha Jariwala\IEEEauthorrefmark{1}, 
Jason J. Qin\IEEEauthorrefmark{1},
Ishaan S. Gupte\IEEEauthorrefmark{1},\\
Bridget Meier\IEEEauthorrefmark{1},
Guanyue Qian\IEEEauthorrefmark{1},
Daniel Abraham\IEEEauthorrefmark{1},
Theodore S. Rappaport\IEEEauthorrefmark{1},
Andreas F. Molisch\IEEEauthorrefmark{2}
}

\IEEEauthorblockA{\IEEEauthorrefmark{1}New York University, Tandon School of Engineering, Brooklyn, NY, USA} 
\IEEEauthorblockA{\IEEEauthorrefmark{2}University of Southern California, Los Angeles, CA, USA}}

\maketitle

\thispagestyle{firststyle}

\begin{abstract}

The necessity of new spectrum for 6G has intensified global interest in radio propagation measurements across emerging frequency bands, use cases, and antenna types. These measurements are vital for understanding the fundamentals of radio channel properties in diverse environments, and involve time-consuming, labor-intensive, and expensive campaigns. A major challenge for the effective utilization of the data generated from propagation measurement campaigns has been the lack of a standardized format for reporting and archiving results. Although organizations such as NIST, NGA, and 3GPP have made commendable efforts toward data pooling, a unified machine-readable data format for consolidating measurements across different institutions and frequencies remains a critical missing piece in advancing global standardization efforts and maximizing the utility of collected data. This paper introduces a new, standardized point-data format for wireless propagation measurements and demonstrates how multiple institutions may merge disparate measurement campaigns into a common format for global access. This data format, when used alongside an environmental map and a measurement summary metadata table, enables the seamless integration of radio propagation data from disparate sources by using a structured representation of key propagation parameters. Here, we show the efficacy of the proposed point-data format standard using data gathered from two independent sub-THz urban microcell (UMi) campaigns: 142 GHz measurements at New York University (NYU) and 145 GHz measurements at the University of Southern California (USC). A joint path loss analysis using the close-in path loss model with a 1 m reference distance yields a refined estimate of the path loss exponent (PLE) employing the proposed standard to pool measurements from the two different organizations. Other key statistics such as RMS delay spread and angular spread are determined using datasets from the two different institutions cast in the standardized format. Adopting this simple, unified format will accelerate comprehensive channel model development, build multi-institutional datasets, and feed AI/ML applications with reliable training data in a common format from many sources.


\end{abstract}


 \section{Introduction}
The wireless telecommunications industry has made significant progress in standardizing and adopting 5G technologies, with 6G standardization efforts set to begin in 2025\cite{Rapp2019ia, Tataria2021pieee}. The radio channel fundamentally dictates the capability of any wireless system. Emerging deployment scenarios (e.g., 3D network topologies, vehicular systems), new applications (e.g., ICAS), and expanded frequency ranges (upper mid-band\cite{Shakya2024ojcoms}, sub-THz\cite{Xing2021icl,abbasi2022thz}) demand both the creation of new channel models and the refinement of existing ones\cite{Shafi2017jsac, Rapp2017tap}. Enormous effort is required to build instrumentation and carry out measurement campaigns for such new applications, scenarios and frequency ranges, with research institutions worldwide engaged in their own efforts to empirically validate new models. However, global harmonization and machine learning–driven modeling efforts to form channel models informed by empirical data are hindered by the absence of an agreed-upon data format for empirical channel measurements. Given the enormous cost of high quality field measurements, the lack of a standardized data representation is wasteful in many respects \cite{Rapp2025icc}.

Most published empirical channel data are presented as aggregated statistics, such as least-squares path loss fits or cumulative distribution functions (CDFs) of delay spread or angular spread. This approach has two major limitations: (i) arduous combination of results from different campaigns (since the average of two path loss fits is not equivalent to the fit over combined datasets), and (ii) inaccessible original data, as knowledge of channel responses at site-specific locations are hidden in CDFs and data pooling from multiple sources is not possible. The combined datasets with explicit representations of site-specific information for each measurement are vital for the creation of a larger sample size which yields more accurate statistics and enables data-driven AI applications\cite{Rapp2025icc}. The data scarcity, resulting from a lack of a standardized empirical data format, slows the development and adoption of AI for wireless research where training on small, inaccurate, or synthetic datasets can introduce unwanted biases in AI models. Only rarely have papers presented point-data in tabular or graphical form (e.g., \cite{Wu2017tap,Mac2015ia2,Mac2015ia}), and when they do, datasets are typically small with limited parameters, and stored in non-uniform formats. In addition, critical information about the measurement system, data processing methods, and environment is often not provided in machine-readable form.

In this paper, two venerable universities with expert empirical propagation teams have pooled propagation measurement data at sub-THz frequencies in a common point-data format for the first time. The simple yet effective point-data table for presenting location-specific propagation measurements can be continually expanded and used for rapid and accurate statistical analysis of propagation behavior from multiple contributing sources by easily accumulating data along any column of the point-data table. The fundamental idea for a point-data format was first explored in \cite{Rapp2011vtm}, and a first point-data format was developed by NYU WIRELESS\cite{Rapp2025icc}. This paper provides several important extensions to the initial point-data format: (i) a measurement summary metadata table is specified, providing all the measurement campaign specific details that are necessary for interpretation of the point-data; (ii) several condensed parameters are defined in the metadata table that are useful for a variety of system evaluations; and (iii) example point-data tables are presented for the sub-THz urban wireless channels measured by NYU and USC reported initially in \cite{Shakya2024tap} and \cite{abbasi2023thz}, respectively. Furthermore, we validate the format reciprocally: both teams analyze the other's data and processing, and we perform joint analysis on the pooled dataset, demonstrating that the point-data format supports large, multiparty contributions within standardization bodies and AI-focused data pools.


The remainder of the paper is organized as follows: Section II explores the data pooling and analysis methods adopted by different standardization bodies. Section III summarizes the key benefits of the point-data format. Section IV introduces metadata tables that describe the channel measurement system and environment. Point-data tables generated from the measurements in the 140 GHz band, conducted independently by USC and NYU. Path loss analysis results using pooled point-data tables from both USC and NYU are presented in Section V before the conclusion.

\section{Current Standardization Mechanisms for Propagation Datasets} 
\subsection{3GPP}

3GPP is a global standardization body focused on the creation of the mobile broadband standard for telecommunications. For the evaluation of different system proposals, it has developed a standardized {\em statistical} channel model, generally called 38.901 \cite{3GPPTR38901}. 3GPP is currently discussing various extensions and modifications of this model, e.g., for the upper mid-band (FR3), as well as definition of an AI-based channel model. However, in all of these activities, contributions from different participants present measurement data only as CDFs with brief descriptions of the measurement setup and environment \cite{r12502381,r12403925} or as tables listing mean, variance, or relevant distribution parameters \cite{r12501169,r12501281}, even though some contributions (e.g., R1-2405865\cite{r12405865}) highlight the difficulty of harmonizing these results due to nonuniform data formats and missing metadata. 
As a result, aggregation of the results to define channel model parameters cannot be done, and the enormous time and expense of the many measurement campaign contributions are wasted due to a lack of a common data format for data pooling, such as proposed herein\cite{Sun2016tvt}.

\subsection{NGA}
The Next-G Alliance (NGA) is a consortium of North American wireless institutions and industry corporations focused on the advancement of wireless technologies through private-sector efforts. NGA technical reports provide a rich collection of channel measurement and modeling insights and results from member institutions. However, similar to 3GPP, the valuable results are disaggregated between institutions and cannot be pooled for collective analysis since no collective data format exists.

\subsection{NIST} 
The NextG Channel Model Alliance, spearheaded by the National Institute of Standards and Technology (NIST), provides an important forum for exchange of measurement data between participants, including storage resources for exchange of raw measurements that span an extensive range of frequencies, from sub-6 GHz bands to the millimeter-wave (mmWave) and sub-Terahertz (sub-THz) spectrums, including 28 GHz, 60 GHz, 73 GHz, 140 GHz, and 300 GHz. These measurements explore diverse environments encompassing indoor spaces such as offices, factory floors, residential settings, and specialized locations such as data centers and basements, alongside critical outdoor urban and suburban scenarios. However, the data provided by the different participants use incompatible formats and without clear definitions of the measurement equipment being used. Hence, groups aiming to use datasets from multiple contributors must analyze those formats and reshape the data in order to enable merging and/or comparison. Such an exercise is costly, time consuming, and prone to introducing error into the original data sets.

\section{Point-Data Format Attributes \& Benefits} 

The point-data format provides a standardized, extensible structure for capturing wireless channel measurements in a site-specific, geometry-aware manner\cite{Rapp2025icc}. Each row in a point-data table corresponds to a unique transmitter–receiver (TX–RX) location, which, importantly, is tied to a site specific map of the environment. Rows can be appended to the table for each new TX-RX measurement location pair in a specific environment, allowing pooling of measurement results and an ever-growing dataset. A measurement summary metadata table (detailed in Section IV) includes spatial metadata, environmental descriptors, and associated channel characteristics such as path loss, RMS delay spread, angular spreads, and antenna polarization. This granularity preserves the physical context of each measurement, which is critical for understanding spatial variation in propagation phenomena. The impact of the specific environment and the variations from location to location are important at any frequency, but are particularly pronounced at high frequencies \cite{Rapp2012rws}. The point-data representation, therefore, enables the creation of expansive robust multi-institutional datasets vital for AI/ML driven modeling and applications.

Unlike traditional aggregated summaries or isolated plots, point-data tables offer a dual advantage: they allow researchers to recreate conventional statistical models (such as CI and ABG---Alpha Beta Gamma---path loss models, empirical delay/angular spread distributions, and CDFs) while simultaneously enabling fine-grained, location-dependent site-specific analysis. This duality makes the format well-suited for both conventional channel modeling and next-generation applications, including training and testing of machine learning-based propagation prediction, and calibration of ray tracing simulations. Furthermore, tabulated measurement-system metadata, including system bandwidth, carrier frequency, antenna beamwidths, dynamic range, and threshold settings, provide essential, transparent context for interpreting the measurements and their environment. 

Using the point-data format, global institutions can contribute to or build upon shared datasets with confidence in the consistency of the underlying measurements. As datasets collected in the point-data format grow, they will support large-scale, multi-site model development across multiple environments, frequencies, and deployment scenarios, and facilitate development of wireless standards as discussed in Section II. 


\section{Measurement Summary Metadata tables for Point-Data Format} 
Measurement summary metadata tables must accompany the point-data tables to provide details about the measurement system employed for the channel measurements. The metadata table lists the essential information for clarity in the measurement and post-processing methodology, including thresholds in the temporal and spatial domains. Measurement summary metadata tables are vital to determine whether datasets are suitable for pooling.

\begin{itemize}[leftmargin=*]
	\item \textbf{Environment (Env.):} The physical scenario where measurements were conducted: UMi, UMa, RMa, InH, and InF.
	\item \textbf{Az and/or El Resolution ($\Delta\phi/\Delta\theta$):} The step size between discrete antenna pointing directions during azimuth/elevation sweeps/cuts, in degrees.
	\item \textbf{Mobility Conditions ($v$):} If any nodes are moving: speed (m/s) and trajectory (coordinate set, e.g., $(x, y, t)$).
	\item \textbf{Frequency ($f_c$):} Start or center frequency (must be specified), provided in MHz.
	\item \textbf{Bandwidth ($BW$):} null-to-null, provided in MHz or GHz.
	\item \textbf{Average TX Power ($P_{TX,avg}$):} Average power fed into the TX Antenna connector, provided in dBm.
	\item \textbf{Max. Dynamic Range ($DR_{\max}$) Or Noise Figure ($NF$) / Receiver Sensitivity:} Max. Dynamic Range is based on the maximum and noise floor per power delay profile (PDP). Noise Figure provided in dB; Receiver Sensitivity provided in dBm.
	\item \textbf{Thresholding per Sample PDP ($T_{\mathrm{PDP}}$):} The delay domain threshold used. e.g., NYU measurements use the greater between 25 dB below the max. PDP power or 5 dB above the noise floor for each PDP. It is fixed for the entire measurement.
	\item \textbf{Thresholding per Sample in Angular Domain ($T_{\mathrm{PAS}}$):} The spatial domain thresholding per beam/spatial lobe. It is fixed for the entire measurement.
	\item \textbf{Maximum Measurable Propagation Delay ($\tau_{\max}$);  Measurement Repetition Rate ($f_{\mathrm{rep}}$):} The maximum measurable time delay and repetition rate.
	\item \textbf{Waveform and averaging details ($L_{\mathrm{PN}},\ N_{\mathrm{avg}}$):} Description of the sounding waveform, including type, length, Peak-to-Average Power Ratio (PAPR), and spreading factor.
	\item \textbf{Sampling time resolution ($\Delta t_s$):} The minimum successive time sample in a PDP, provided in ns.
	\item \textbf{Sampling rate ($f_s$):} ADC sampling frequency used during acquisition, reported in MSps.
	\item \textbf{Synchronization Details ($\mathrm{Sync.}$):} Description of the synchronization using one among reference clock, GPS sync, PTP-based\cite{Shakya2023gc}, or an external trigger sync algorithm.
	\item \textbf{Sweep Parameters Frequency Domain ($\mathrm{Sweep\ Params\ (FD)}$):} List of the frequency domain sweep parameters, including Intermediate Frequency Bandwidth (IFBW), Number of points per sweep ($N_{\mathrm{pts}}$), and Averaging.
	\item \textbf{AS Definition ($\mathrm{AS~Def.}$):} Fleury's or 3GPP definition.
	\item \textbf{Antenna Model Number ($\mathrm{Ant.~Model}$) \& Operational Frequency ($f_{Ant,op}$):} Model number provided by the manufacturer and the specified operating frequency range.
	\item \textbf{TX/RX Antenna Type ($\mathrm{Ant.~Type}$):} Horn, dipole, or patch array.
	\item \textbf{Antenna Bandwidth ($BW_{Ant.}$):} Antenna bandwidth specified by manufacturer, provided in GHz for $S_{11} < -10$ dB. May not necessarily match system bandwidth.
	\item \textbf{TX/RX Antenna Gain on Boresight ($G_{TX} / G_{RX}$):} The manufacturer specified or measured antenna gain in the boresight direction, provided in dBi.
	\item \textbf{TX/RX Antenna HPBW $\theta_{3\mathrm{dB,TX}}/ \theta_{3\mathrm{dB,RX}}$:} Half Power Beamwidth (HPBW) of the antenna, manufacturer specified, provided in degrees.
	\item \textbf{Sidelobe Level ($\mathrm{SLL}$):} Maximum, worst-case sidelobe level w.r.t main lobe peak measured at boresight, provided in dB.
	\item \textbf{Front-to-Back Ratio ($\mathrm{FBR}$):} Ratio of the gain in the maximal direction to the gain in the opposite direction 180 $^\circ$ away, provided in dB.
	\item \textbf{XPD:} Ratio of power received in co-polarization configuration to the power received in cross-polarization configuration, provided in dB\cite{Xing2018vtc}.
	\item \textbf{Polarization Type ($\mathrm{Pol.}$):} Specification of the electric field polarization radiated from the antenna, linear, circular, or dual polarization.
	\item \textbf{Array Geometry:} Antenna arrays geometry: Uniform Linear Array (ULA), Uniform Planar Array (UPA). Spacing between antenna elements (provided in mm).
	\item \textbf{Number of Elements:} The number of elements used in the antenna array.
\end{itemize}

Table \ref{tab:preCompare} shows a measurement-summary metadata for two systems: the NYU WIRELESS sliding-correlation, time-domain channel sounder \cite{Shakya2023gc} and the RF-over-fiber, frequency-domain channel sounder at USC \cite{abbasi2023thz}.
\begin{table}[htbp]
	\caption{Comparison of measurement summary metadata tables for NYU\cite{Shakya2023gc} and USC\cite{abbasi2023thz} UMi campaigns.}
	\label{tab:preCompare}
	\centering
	\begin{adjustbox}{width=9cm}
		\begin{tabular}{| >{\centering\arraybackslash}p{2.3cm}
				| >{\centering\arraybackslash}p{3.05cm}
				| >{\centering\arraybackslash}p{3.05cm}|}
			\hline
			\textbf{Parameter} & \textbf{NYU\cite{Shakya2024tap}} & \textbf{USC\cite{abbasi2023thz}}\\
			\hline
			$\mathrm{Env.}$ & UMi & UMi \\
			\hline
			$\Delta\phi/\Delta\theta$ & 8$^\circ$/8$^\circ$ & 10$^\circ$/13$^\circ$ \\
			\hline
			$v$ & Static & Static \\
			\hline
			$f_c $& 142 GHz (center) & 145.5 GHz (center) \\
			\hline
			$BW$ & 1 GHz & 1 GHz \\
			\hline
			$P_{\mathrm{TX,avg}}$ & 0 dBm & -1 dBm \\
			\hline
			$DR_{\max}$ & 40 dB & 40 dB \\
			\hline
			$ T_{\mathrm{PDP}} $ & max(25 dB below peak, 5 dB above noise floor) & $ \tau_{\mathrm{gate}} = 966.67 $ ns; $ +12 $ dB (noise) \\
			\hline
			$ T_{\mathrm{PAS}} $ & 10 dB below max. PAS power & $ \tau_{\mathrm{gate}} = 966.67 $ ns; $ +12 $ dB (noise) \\
			\hline
			$ \tau_{\max};\ f_{\mathrm{rep}} $ & 4094 ns; 32.752 ms & 1 $ \mu $s; -- \\
			\hline
			$ L_{\mathrm{PN}},\ N_{\mathrm{avg}} $ & 2047 chip PN; Sliding corr.; 20 PDP avg. & --; --; No averaging \\
			\hline
			$ \Delta t_s $ & 1 ns & 1 ns \\
			\hline
			$ f_s $ & 2.5 Msps & -- \\
			\hline
			$ \mathrm{Sync.} $ & Rubidium clocks at TX \& RX & VNA Internal \\
			\hline
			$\mathrm{Sweep\ Params}$ & -- & IFBW 10 kHz; $ N_{\mathrm{pts}} = 1001 $\\
			\hline
			$ \mathrm{AS\ Def.} $ & 3GPP TR 38.901 & Fleury \\
			\hline
			$ \mathrm{Ant.\ Model};\ f_{Ant,op} $ & Mi-Wave 261D-27; D-band & VDI WR-5.1; G-band \\
			\hline
			$ \mathrm{Ant.\ Type} $ & Pyramidal horn & Conical horn\\
			\hline
			$ BW_{Ant.} $ & 60 GHz & 80 GHz \\
			\hline
			$ G_{TX}/G_{RX} $ & 27 dBi & 21 dBi \\
			\hline
			$ \theta_{3\mathrm{dB,TX}}/\theta_{3\mathrm{dB,RX}} $ & 8$^\circ$ & 13$^\circ$ \\
			\hline
			$ \mathrm{SLL} $ & -11 dB & -13 dB \\
			\hline
			$ \mathrm{FBR} $ & 30 dB & 35 dB \\
			\hline
			$ \mathrm{XPD} $ & 29.2 dB \cite{Xing2018vtc} & -- \\
			\hline
			$ \mathrm{Pol.} $ & Linear & Linear \\
			\hline
		\end{tabular}
	\end{adjustbox}
\end{table}

\vspace{-10 pt}

\section{NYU and USC Measurements in Point-Data Format}
The point-data tables presented here for the NYU and USC sub-THz UMi measurements follow the format established in \cite{Rapp2025icc}. The columns of the point-data table summarize the large-scale spatio-temporal statistics at each valid TX-RX location pair measured in a given environment for both NYU and USC. 

\begin{figure}[htbp]
	\centering%
	\includegraphics[width=0.92\columnwidth]{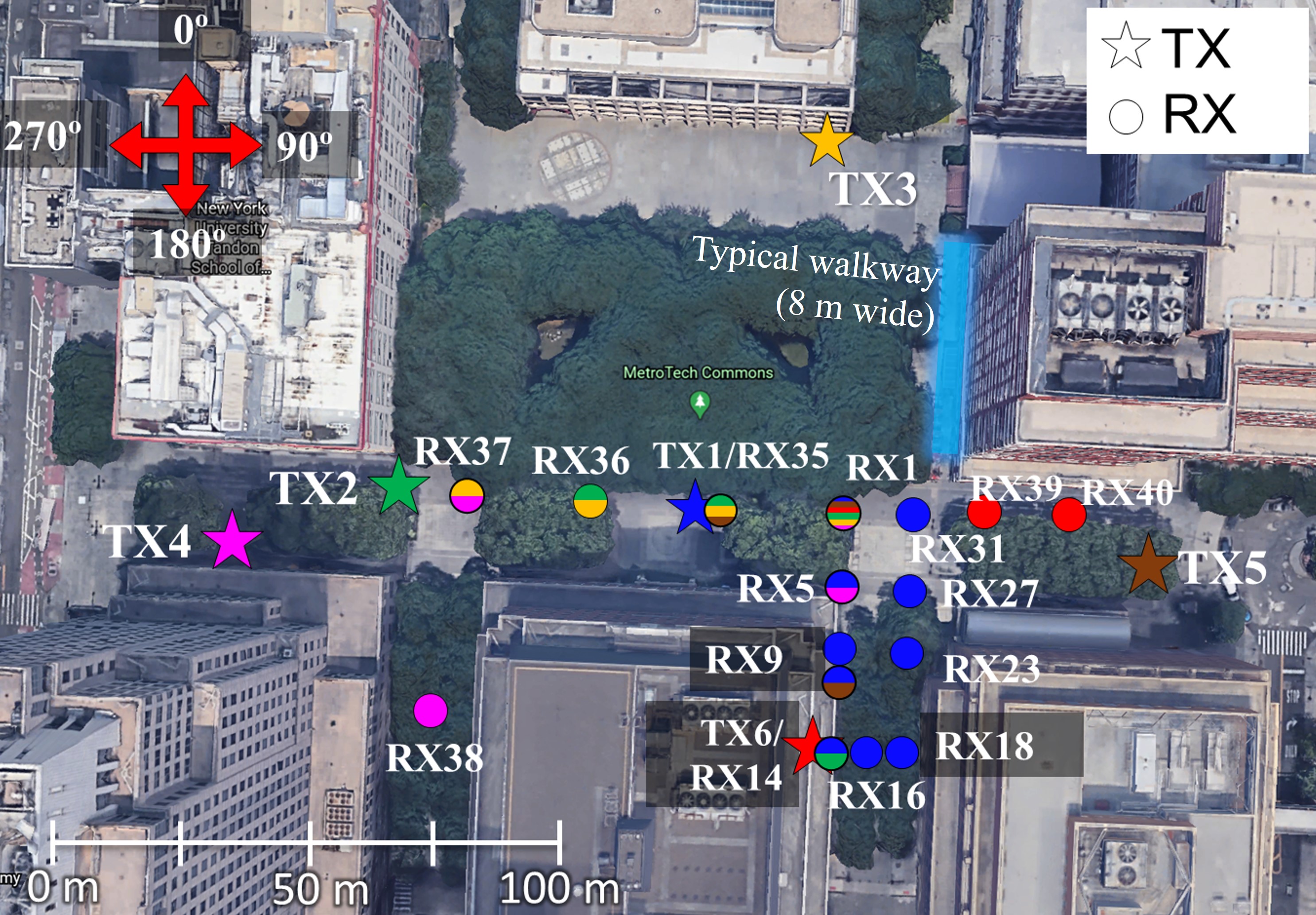}
	\caption{Map of the environment for the NYU WIRELESS 142 GHz propagation measurements in Brooklyn, New York. 27 TX-RX locations are measured with 16 LOS and 11 NLOS \cite{Shakya2024tap}.}
	\label{fig:NYU140UMi}
	\vspace{-10pt}
\end{figure}

An example of the point-data table encompassing the NYU WIRELESS 142 GHz UMi propagation measurements conducted in MetroTech Commons near the NYU campus in Downtown Brooklyn, New York is presented in Table \ref{tab:PDT142NYU}. Six (3 LOS and 3 NLOS) out of the 27 locations measured in the NYU 142 GHz campaign are shown in Table \ref{tab:PDT142NYU}. The measurement environment showing all 27 TX-RX location pairs (16 LOS, and 11 NLOS) is presented in Fig. \ref{fig:NYU140UMi}. The measured open square at NYU is a unique urban propagation environment consisting of buildings, orchards with walkways, lampposts, benches, sculptures, and several pedestrians.

\begin{figure}[htbp]
	\centering%
	\includegraphics[width=0.94\columnwidth]{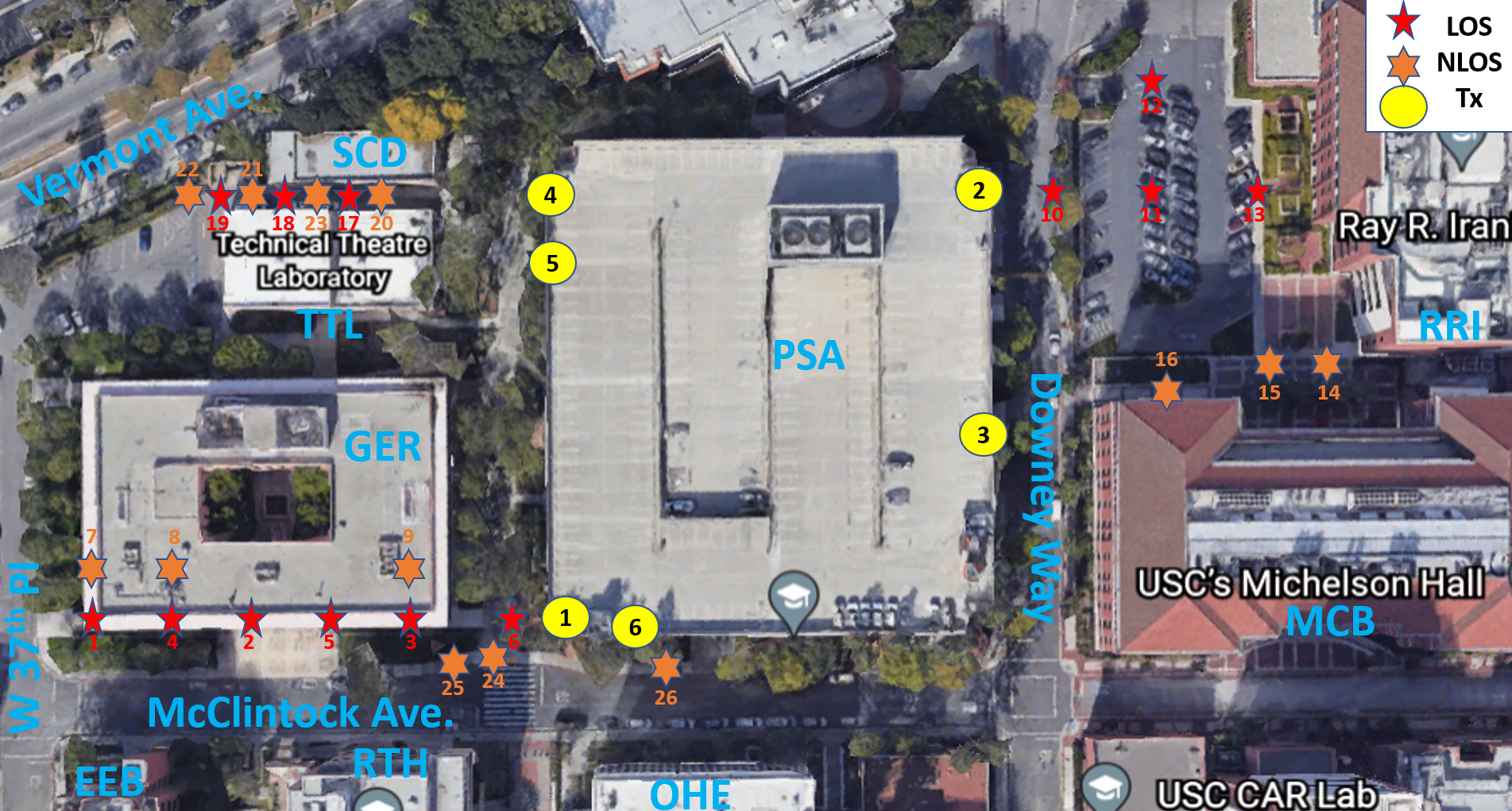}
	\caption{Map of the USC 145 GHz propagation measurement campaign conducted in Los Angeles, CA. A total of 26 TX–RX location pairs were measured, comprising 13 LOS and 13 NLOS links \cite{abbasi2022thz}.}
	\label{fig:USC140}
\end{figure}

Similarly, Table \ref{tab:PDT145USC} presents point-data from USC's 145 GHz UMi campaign in Los Angeles, CA, listing six representative TX–RX pairs (three LOS, three NLOS); subsequent sections analyze the full set of twenty-six pairs. The USC measurement environment is depicted in Fig. \ref{fig:USC140}. The environment comprises of office buildings, parking structures, and landscaped areas with vegetation, featuring both open spaces and locations partially shaded or obstructed by architectural elements and foliage~\cite{abbasi2022thz,abbasi2023thz}.

\renewcommand{\arraystretch}{1.2}
\begin{table*}[htbp]
	\centering
	\color{black}
	\caption{Example point-data table generated from the NYU WIRELESS measurements in Brooklyn, NY, showing site-specific large-scale spatio-temporal statistics. Fig. \ref{fig:NYU140UMi} shows the corresponding map of the measured environment \cite{Shakya2024tap}. }
	\vspace{-0.1cm}
	\begin{tabular}{p{0.6 cm}p{0.5 cm}p{0.5 cm}p{0.69 cm}p{0.6 cm}p{0.69 cm}p{0.69 cm}p{0.69 cm}p{0.69 cm}p{0.69 cm}p{0.69 cm}p{0.69 cm}p{0.6 cm}p{0.6 cm}p{0.6 cm}p{0.6 cm}}
		\hline
		\multicolumn{1}{p{0.6 cm}}{\textbf{Freq.}} & \textbf{TX} & \textbf{RX} & \textbf{Loc.} & \multicolumn{1}{p{0.6 cm}}{\textbf{TR Sep.}}  & \multicolumn{1}{p{0.69 cm}}{\textbf{ PL}} & \multicolumn{1}{p{0.69 cm}}{\textbf{Mean Dir. DS}} & \multicolumn{1}{p{0.69 cm}}{\textbf{Omni DS}} & \multicolumn{1}{p{0.69 cm}}{\textbf{Mean Lobe ASA}} & \multicolumn{1}{p{0.69 cm}}{\textbf{Omni ASA}} & \multicolumn{1}{p{0.69 cm}}{\textbf{Mean Lobe ASD}} & \multicolumn{1}{p{0.69 cm}}{\textbf{Omni ASD}} & \multicolumn{1}{p{0.6 cm}}{\textbf{Mean Lobe ZSA}} & \multicolumn{1}{p{0.6 cm}}{\textbf{Omni ZSA}} & \multicolumn{1}{p{0.6 cm}}{\textbf{Mean Lobe ZSD}} & \multicolumn{1}{p{0.6 cm}}{\textbf{Omni ZSD}} \\
		\hline
		\text{[GHz]}& & & &[m]&[dB]&[ns]&[ns]&[$^\circ$]&[$^\circ$]&[$^\circ$]&[$^\circ$]&[$^\circ$]&[$^\circ$]&[$^\circ$]&[$^\circ$] \\ 
		\hline
		\multirow{6}{*}{\textbf{142}} & \multirow{3}{*}{TX1}& RX1& LOS& 24.43& 102.6& 50.8& 15.7& 2.3& 21.2& 2.8& 20.1& 3.1& 5.4& 3.2& 3.3
		\\
		&       & RX5& LOS& 27.22& 102.1& 0.1& 6.1& 2.5& 2.5& 2.5& 4.4& 3.3& 6.6& 3.3& 3.4
		\\
		&       & RX9& NLOS& 32.65& 126.5& 9.0& 19.1& 3.9& 68.8& 2.5& 2.5& 3.7& 6.3& 3.4& 3.4
		\\
		\cline{2-16}
		& \multirow{3}{*}{TX6} & RX1& LOS& 39.08& 105.6& 2.2& 101.0& 3.0& 3.0& 3.7& 11.4& 3.8& 5.4& 1.2& 1.1
		\\
		&       & RX39& NLOS& 48.65& 114.7& 1.7& 23.9& 3.4& 61.8& 4.2& 32.0& 3.0& 6.0& 1.8& 1.7
		\\
		&       & RX40& NLOS& 53.02& 121.7& 1.9& 29.8& 3.4& 34.0& 3.3& 19.8& 3.8& 5.4& 2.2& 2.1\\
		\hline
	\end{tabular}%
	\label{tab:PDT142NYU}%
\end{table*}%
\vspace{-5 pt}
\renewcommand{\arraystretch}{1.0}

\renewcommand{\arraystretch}{1.2}
\begin{table*}[htbp]
	\centering
	\color{black}
	\caption{Example point-data table generated from the USC's measurements in Los Angeles, CA, showing site-specific large-scale spatio-temporal statistics. Fig. \ref{fig:USC140} shows the corresponding map of the measured environment \cite{abbasi2023thz}.}
	\vspace{-0.1cm}
	\begin{tabular}{p{0.6 cm}p{0.5 cm}p{0.5 cm}p{0.69 cm}p{0.6 cm}p{0.69 cm}p{0.69 cm}p{0.69 cm}p{0.69 cm}p{0.69 cm}p{0.69 cm}p{0.69 cm}p{0.6 cm}p{0.6 cm}p{0.6 cm}p{0.6 cm}}
		\hline
		\multicolumn{1}{p{0.6 cm}}{\textbf{Freq.}} & \textbf{TX} & \textbf{RX} & \textbf{Loc.} & \multicolumn{1}{p{0.6 cm}}{\textbf{TR Sep.}}  & \multicolumn{1}{p{0.69 cm}}{\textbf{ PL}} & \multicolumn{1}{p{0.69 cm}}{\textbf{Mean Dir. DS}} & \multicolumn{1}{p{0.69 cm}}{\textbf{Omni DS}} & \multicolumn{1}{p{0.69 cm}}{\textbf{Mean Lobe ASA}} & \multicolumn{1}{p{0.69 cm}}{\textbf{Omni ASA}} & \multicolumn{1}{p{0.69 cm}}{\textbf{Mean Lobe ASD}} & \multicolumn{1}{p{0.69 cm}}{\textbf{Omni ASD}} & \multicolumn{1}{p{0.6 cm}}{\textbf{Mean Lobe ZSA}} & \multicolumn{1}{p{0.6 cm}}{\textbf{Omni ZSA}} & \multicolumn{1}{p{0.6 cm}}{\textbf{Mean Lobe ZSD}} & \multicolumn{1}{p{0.6 cm}}{\textbf{Omni ZSD}} \\
		\hline
		\text{[GHz]}& & & &[m]&[dB]&[ns]&[ns]&[$^\circ$]&[$^\circ$]&[$^\circ$]&[$^\circ$]&[$^\circ$]&[$^\circ$]&[$^\circ$]&[$^\circ$] \\ 
		\hline
		\multirow{6}{*}{\textbf{145}} & \multirow{6}{*}{TX1}& RX1& LOS& 82.5 & 111.9 &	4.2 & 48.4 & 7.8 & 43.3 & 11.5 & 20.5 & 7.7 & 8.6 & 7.9 & 8.3
		\\
		&       & RX4& LOS& 72.3 & 109.8 & 2.4 & 67.1 & 7.8 & 30.1 & 6.7 & 14.7	& 6.7 & 7.1 &	6.5 & 6.9
		\\
		&       & RX5& LOS& 49.8& 107.7 & 1.7 & 28.9 & 6.8 & 14.8 & 7.1 & 11.5 & 6.5& 6.8 & 6.7 & 7.0
		\\
		&  & RX7& NLOS& 83 & 130.0 & 117.6 & 121.6 & 121.1 & 121.1 & 38.4 & 38.4 & 10.5 & 10.5 & 10.0 & 10.0
		\\
		&       & RX8& NLOS& 73 & 130.2 & 88.1 & 97.7 & 101.2 & 101.1 & 36.9 & 36.9 & 10.3 & 10.3 & 10.0 & 10.0
		\\
		&       & RX9& NLOS& 46 & 124.7 & 36.2 & 67.7 & 73.3 & 81.4 & 40.6 & 40.5 & 9.9 & 10.0 & 10.2 & 10.2
		\\
		\hline
	\end{tabular}%
	\begin{flushleft}
		\footnotesize{Note: The point-data format allows for the concatenation of Tables \ref{tab:PDT142NYU} and \ref{tab:PDT145USC} to form a larger homogeneous data set of pooled measurements, by which the statistics can be computed down the columns of the combined tables.}
	\end{flushleft}
	\vspace{-15pt}
	\label{tab:PDT145USC}%
\end{table*}%
\vspace{-5 pt}
\renewcommand{\arraystretch}{1.0}

\section{Analysis using the Point-Data Format} 
This section demonstrates how the standardized point-data format enables seamless combination of independent measurement campaigns. The combined NYU and USC dataset consists of 53 TX–RX location pairs, 27 measured by NYU and 26 measured by USC. While only six representative locations from each campaign are tabulated in this paper due to space constraints, the analysis of the path loss and delay spread, presented here, uses all 53 checked and validated measurement points. 

\begin{figure}[htbp]
	\centering%
	\vspace{-10 pt}
	\includegraphics[width=0.98\columnwidth]{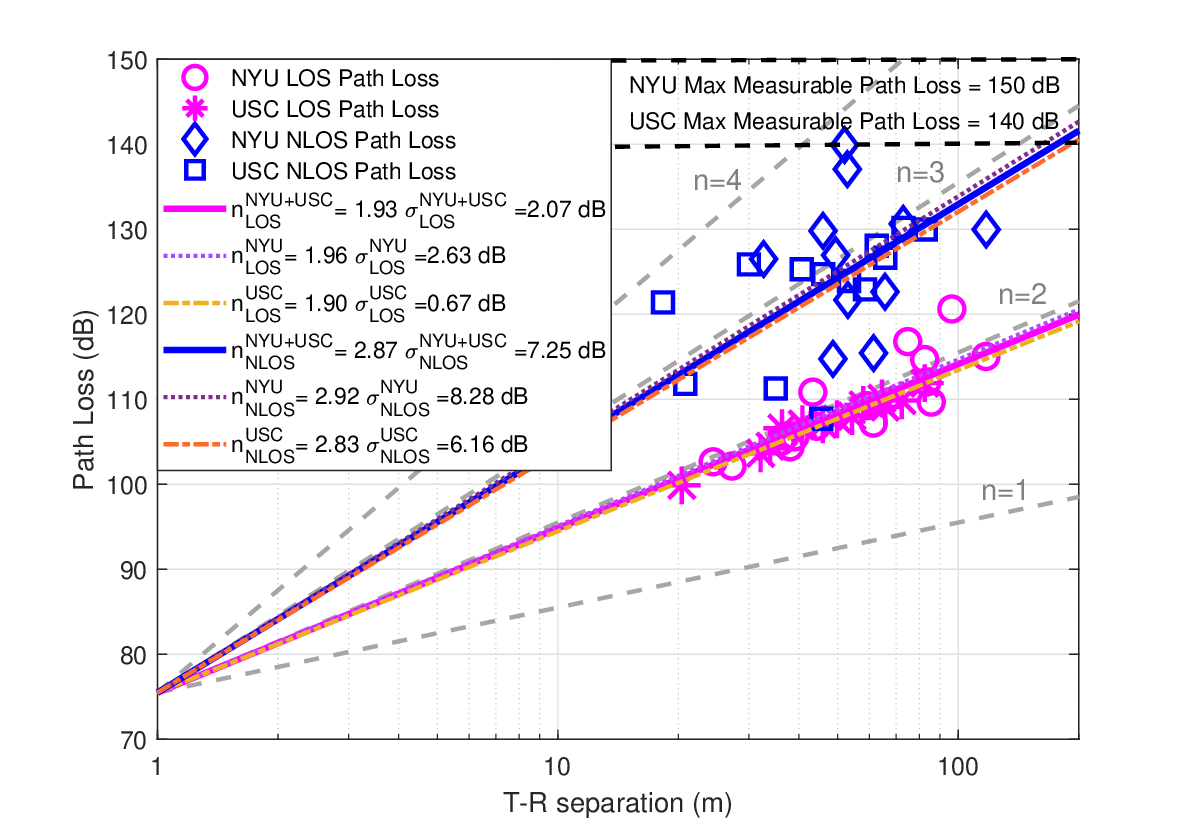}
	\caption{CI PL scatter plot and best PLE fit combining NYU and USC data across 53 locations.}
	\label{fig:CIplot}
	\vspace{-15pt}
\end{figure}

A path loss scatter plot for the combined dataset is shown in Fig. \ref{fig:CIplot}, along with best-fit curves obtained using the CI path loss model with a 1 m reference distance. Distinct LOS and NLOS clusters are clearly visible, with LOS measurements from both sites exhibiting lower path loss than NLOS. The best-fit PLE values for the merged dataset (1.93 LOS, 2.87 NLOS) closely match those derived from the individual campaigns, confirming that the point-data format enables accurate merging without distorting underlying propagation trends. 

\begin{figure}[htbp]
	\centering%
	\includegraphics[width=0.94\columnwidth]{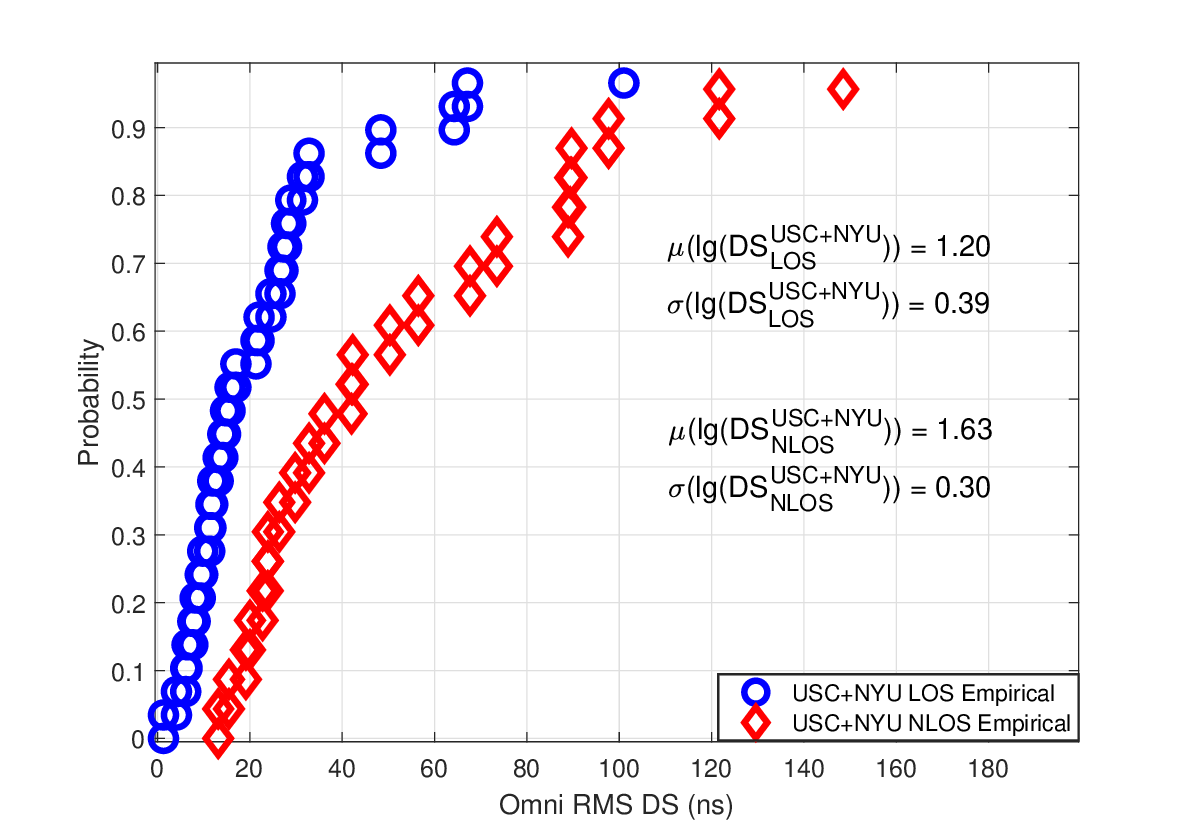}
	\caption{Omnidirectional RMS DS CDF generated from combined USC and NYU datasets spanning 53 TX-RX locations.}
	\label{fig:DSplot}
	\vspace{-15pt}
\end{figure}

The CDFs of the omnidirectional RMS delay spreads for the combined datasets are presented in Fig. \ref{fig:DSplot}. The LOS CDFs from both environments are tightly clustered with low mean delay spreads of 19.7 ns, suggesting limited multipath dispersion in unobstructed links. In contrast, the NLOS CDFs exhibit broader spreads for both datasets. 
Combined, the mean NLOS RMS DS is obtained as 47.3 ns, assuming an underlying log-normal distribution. Notably, the CDFs derived from the merged dataset align closely with those from the individual environments, while reducing ambiguity of the underlying distribution for the urban environment. The joint CDFs, therefore, illustrate that the standardized point-data format retains distinct environmental characteristics, while enabling unified statistical analysis. Table \ref{tab:PLresults} summarizes the LOS and NLOS PLEs, standard deviations, and mean omnidirectional RMS delay spreads for NYU-only, USC-only, and merged datasets, providing a clear side-by-side comparison that highlights the benefits of pooled analysis.

\renewcommand{\arraystretch}{1.3} 
\begin{table}[htbp]
	\centering
	\caption{CI path loss model parameters using a 1 m free space reference distance, along with mean omnidirectional RMS delay spreads, for NYU-only (27 points), USC-only (26 points), and combined datasets (53 points total).}
	\begin{tabular}{|>{\centering\arraybackslash}m{1.7cm}|
			>{\centering\arraybackslash}m{0.55cm}|
			>{\centering\arraybackslash}m{0.65cm}|
			>{\centering\arraybackslash}m{0.55cm}|
			>{\centering\arraybackslash}m{0.65cm}|
			>{\centering\arraybackslash}m{0.85cm}|
			>{\centering\arraybackslash}m{0.85cm}|}
		\hline
		\multirow{3}{*}{\parbox{1.7cm}{\centering\textbf{Dataset}}} & \multicolumn{4}{c|}{\textbf{CI PL w/ 1 m ref.}} & \multicolumn{2}{c|}{\textbf{Omni RMS DS}} \\
		\cline{2-7}
		& \multicolumn{2}{c|}{LOS} & \multicolumn{2}{c|}{NLOS} & \multicolumn{1}{c|}{LOS} & \multicolumn{1}{c|}{NLOS} \\
		\cline{2-7}
		& PLE & $\sigma$[dB] & PLE & $\sigma$[dB] & $\mathbb{E}(\cdot)$[ns] & $\mathbb{E}(\cdot)$[ns] \\
		\hline
		\parbox{2cm}{\RaggedRight\textbf{NYU Only} \cite{Shakya2024tap} } & 1.96 & 2.63 & 2.92 & 8.28 & 15.4 & 27.3 \\
		\hline
		\parbox{2cm}{\RaggedRight\textbf{USC Only} \cite{9790802}} & 1.90 & 0.67 & 2.83 & 6.16 & 21.6 & 67.5 \\
		\hline
		\parbox{2cm}{\RaggedRight\textbf{USC +NYU}} & 1.93 & 2.07 & 2.87 & 7.25 & 19.7 & 47.3 \\
		\hline
	\end{tabular}%
	\label{tab:PLresults}%
\end{table}%
\vspace{-5 pt}
\renewcommand{\arraystretch}{1.0}

Pooling the 53 TX-RX location dataset provides a broader representation of urban environments with a wide variety of scatterers. Such a broad campaign is not feasible for a single institution; however, the proposed point-data format enables seamless integration and joint analysis of measurements from both campaigns. Despite some differences in environment and deployment geometry, the combined dataset shows strong consistency in path loss and delay spread characteristics, demonstrating the format’s effectiveness in harmonizing diverse measurement campaigns without losing environment-specific features.

\section{Conclusion} 
A review of measurement campaigns across industry reports, repositories, and academic research has highlighted the undeniable absence of a standard data representation for wireless propagation data. A huge waste in time, effort and money exists at a time when AI/ML tools need as much high quality field data as possible. In this paper, we have demonstrated the point-data format, validated and implemented by the research groups at NYU and USC for outdoor radio propagation measurements conducted in the 140 GHz band. The utility of the multidimensional point-data format in bringing together datasets from different institutions with both statistical and site-specific information is highlighted through the point-data tables presented and the joint statistical analysis conducted for both CI path loss exponent and omnidirectional RMS delay spread. The point-data representation is effective for statistical analysis with scatter plots and CDFs; at the same time, it enables the enrichment of available propagation datasets through direct concatenation, vital for data-driven applications using AI/ML techniques. Universal adoption of this format by both industry and academia will enable effective utilization of the radio propagation data collected by institutions through intense measurement campaigns, while accelerating standardization and AI training and modeling efforts leveraging an ever-expanding pool of reliable measurement data. 


\bibliographystyle{IEEEtran}
\bibliography{references} 

\end{document}